\documentclass[12pt,aps,prb,ajp]{revtex4}   

\usepackage{amsmath}    
\usepackage{graphicx}   
\usepackage{verbatim}

\renewcommand{\footnotetext}[1]{\endnotetext{#1}}
\begin{document}


\title{Visualizing Flat Spacetime: \\

    Viewing Optical versus Special Relativistic Effects\footnote{Copyright (2007) American Association of Physics Teachers. This article may be downloaded for personal use only. Any other use requires prior permission of the author and the American Association of Physics Teachers}}

\author{Don V. Black\footnote{Department of Electrical Engineering and Computer Science, UCI; email:dblack@ieee.org}
         \  M. Gopi\footnote{Department of Information and Computer Science, UCI; email:gopi@uci.edu}
         \  F. Wessel\footnote{Department of Physics \& Astronomy, UCI; email:fwessel@uci.edu}
         \  R. Pajarola\footnote{Department of Visualization and Multimedia Lab, University of Z\"{u}rich; email:pajarola@acm.org}
         \  F. Kuester\footnote{California Institute for Telecommunications and Information Technology, UCSD; email:fkuester@ucsd.edu}
       }

\affiliation{\scriptsize {University of California at Irvine}}

\date{October 15, 2006}
\begin{abstract}

A simple visual representation of Minkowski spacetime appropriate
for a student with a background in geometry and algebra is
presented. Minkowski spacetime can be modeled with a Euclidean
4-space to yield accurate visualizations as predicted by special
relativity theory. The contributions of relativistic aberration as
compared to classical pre-relativistic aberration to the geometry
are discussed in the context of its visual representation.

\end{abstract}

\maketitle

\section{Introduction}

This paper presents a Euclidean 4D model that can be used to view
and explain Minkowski spacetime without resort to higher
mathematics. The simple intuitive method presents the fundamental
concepts underlying the Theory of Special Relativity and enables the
teacher to lead a student from Euclidean geometry into flat
spacetime.

For simplicity, \textbf{\emph{temporal homogeneity}}~\cite{temporal}
and a flat~\cite{flat}
spacetime with no acceleration are assumed, and lighting effects are
not considered. Under these conditions flat Minkowski spacetime is
Euclidean for an inertial observer. The corresponding model can then
be viewed and animated based on 4D raytracing.

\emph{\textbf{Temporal extrusion}} of an inertial 3D object into
4-space along its normalized velocity 4-vector (worldline), followed
by the \textbf{\emph{Lorentz transformation}} (length contraction
and time dilation) of the object into the inertial reference frame
of the stationary camera are used to model object behavior. The
camera can then be moved along the time axis, raytracing the 4D
space, and creating an image collection that can subsequently be
composed into a video sequence capturing the time-varying
effects.\cite{DVBlack:2004}

In the following sections we will: discuss our fundamental
assumptions, and the Minkowski 2D and 3D spacetime diagrams;
describe our model, and the construction of 4D objects from 3D
objects; and finally, demonstrate the resulting animations of 3D
objects in 4D spacetime.

\label{sec:intro}
\section{Theory}

As pointed out in this journal,\cite{Deissler:2005} and demonstrated
by Terrell\cite{Terrell:1959} and Penrose,\cite{Penrose:1959} the
visual phenomena we explore here can be described as the combination
of a pre-relativistic purely optical effect due to finite lightspeed
that was discovered by Roemer in 1677,\cite{Roemer:1677} and special
relativity's four dimensional spacetime discovered by Minkowski in
1908.\cite{Minkowski:1908} The finite speed of light leads to
effects analogous to those of sound, as in the case of locating the
position of a fast high flying jet by the sound of its engines.
Finite and invariant lightspeed requires the physical phenomena
predicted by special relativity: time dilation and length
contraction. Time dilation is observable only if there is a
variation in the object during the viewing period, as in the muon
particle's decay. Length contraction is observable by differences in
the geometry of a relativistic object at rest and in motion.

\subsection{Background}
\label{sec:back}

Relativistic 4D spacetime ($t,x,y,z$) consisting of both space and
time, is often labeled (3+1)D, referring to three spatial dimensions
($x,y,z$) and one time dimension $t$. Similarly, a 3D spacetime
($t,x,y$) could be referred to as (2+1)D, which is to say containing
two spatial dimensions ($x,y$) and one time dimension $t$.

\begin{figure}[htb]
  \centering
  \includegraphics[width=1.0\linewidth]{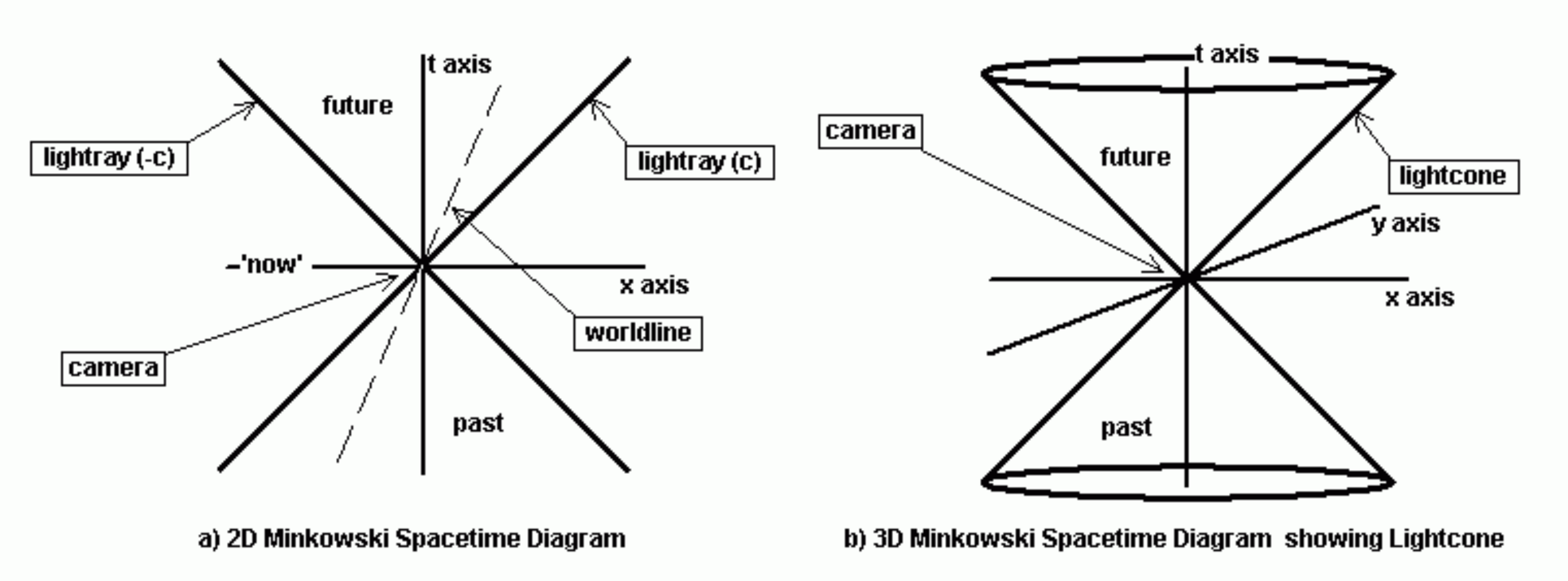}
  \caption{\label{fig:Mink2D}  (1+1)D and (2+1)D Minkowski spacetime diagrams}
  \scriptsize{A camera at the origin can only 'see' an event in the past whose
  lightray passes from that event through the camera at the origin.}
\end{figure}

The most convenient units for the purposes discussed here are
\emph{\textbf{relativistic units}} where $c=1$. The benefit of using
relativistic units is that the units along all the spacetime axes
have the same scale, resulting in a lightray traveling one unit
along the spatial axes for each unit it travels along the time axis.
A lightray $c$ can thus be represented in a Minkowski 2D spacetime
diagram as a $45^{\circ}$ bisector, or in a 3D spacetime diagram as
the surface of a right circular cone, both shown in
Figure~\ref{fig:Mink2D}. We will use the light-second, the distance
light travels in a second, as the basic unit of measure in our
discussion.

An object's \textbf{\emph{worldline}} is its 4D path through
spacetime. The instantaneous direction of an object's worldline is
the object's proper time axis.  The slope of this proper time axis
in the Minkowski diagram represents the object's speed. The
worldline through flat spacetime of an object with a constant
velocity is a straight line. The normalized tangent to an object's
worldline is the object's instantaneous velocity 4-vector.

\begin{figure}[htb]
  \centering
  \includegraphics[width=.9\linewidth]{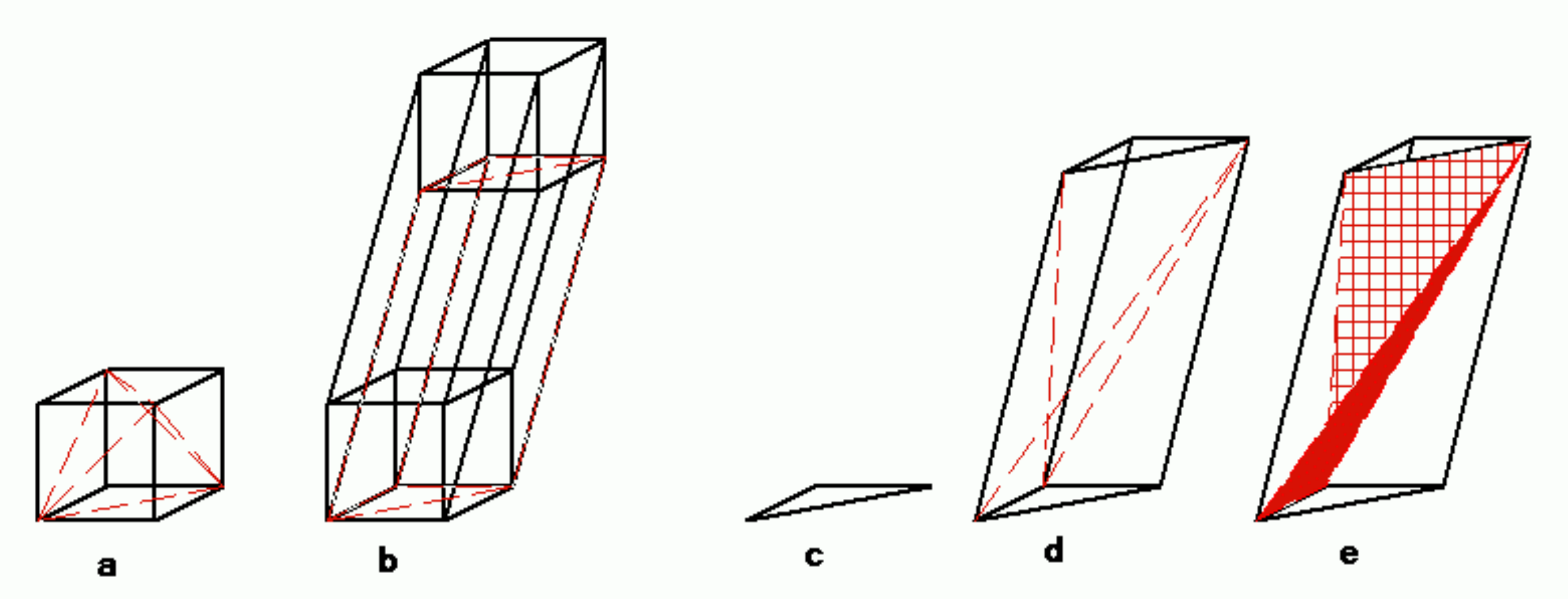}
  \caption{\label{fig:fig322}   Cube \& triangle: Extruded then tessellated}
\end{figure}

A 3D object can be created by
\textbf{\emph{extruding}}~\cite{Extrusion}
a 2D object in a direction perpendicular to the 2D plane in which
the object lies (for example, by extruding a square from the $X,Y$
plane along the $Z$ axis). Likewise, a 4D object can be created by
extruding a 3D object in a direction orthogonal to the 3D hyperplane
in which the object lies. A 4D example would be the extrusion of a
cube from the $X,Y,Z$ 3-space, along the $T$ axis. Two examples are
shown in Figure~\ref{fig:fig322}. We call this
\textbf{\emph{temporal extrusion}} when a 3D object is extruded
along its worldline.

Raytracing\cite{Glassner:1989a} is a geometric 3D image rendering
algorithm that colors the pixel on a viewplane by sending a ray from
the viewpoint, through a pixel on the viewplane, and out into the
scene's 3-space where it may intersect the 2D surface element (such
as a triangle) used to define the boundaries of a 3D object. The
color of the object's surface at the intersection is used to color
the corresponding pixel in the viewplane. This procedure is repeated
for each of the pixels in the viewplane. Howard~\cite{Howard:1995}
adapted the open-source 3D raytracer POV-Ray Version 2.0 to
relativistic raytracing by changing the angle of incidence as a
light ray passes from one inertial reference frame to another. We
found it necessary to increase the model's flexibility in order to
demonstrate the difference between finite lightspeed effects and
relativistic effects.

We have developed a simple four dimensional raytracer by globally
extending a 3D raytracer's\cite{Kirk:1988} vector math from 3D to 4D
and adding a fourth component $t$ to the coordinate system. We
constrained the lightrays to lie on the negative lightcone so the
ray travels through the model at lightspeed. The resulting 4D
raytracer can image a Euclidean 4D space of 4D objects.

It can be shown
that the length of an object with an arbitrary constant
relativistic~\cite{relativisitic}
~velocity $\beta~=~\frac{v}{c}$ will contract in the direction of
motion by a factor of $\frac{1}{\gamma}~=~\sqrt{1-\beta^2}$. It can
also be shown
that the proper duration between any two events on the relativistic
object's worldline will expand (dilate) by the Lorentz factor
$\gamma~=~\frac{1}{\sqrt{1-\beta^2}}$. This is known as
\textbf{\emph{length contraction}} and \textbf{\emph{time
dilation}}, respectively.

\subsection{Object Construction}
\label{sec:cons}

Any 3D object defined by bounding triangles such as the cube in
Figure~\ref{fig:fig322}a can be \textbf{\textsl{temporally
extruded}} into a 4D hyperobject and inserted in the scene's 4-space
by extruding each $f$ of its $n$ individual triangles as follows.
Assuming that the triangle's vertices are defined by their 3D
coordinates in 3-space, insert a $t$ component into each of the
vertex coordinates and set $t$ to some constant value, say $t_{0}$.

\vspace {-0.1in}
\begin{center}
$(x_{i}, y_{i}, z_{i})_{f} \rightarrow (t_{0}, x_{i}, y_{i},
z_{i})_{f}$.
\end{center}
\vspace {-0.1in}

When performed on all three vertices $i$, the 2D triangle $f$ will
have a unique location in 4-space.

\begin{figure}[htb]
  \centering
  \includegraphics[width=.9\linewidth]{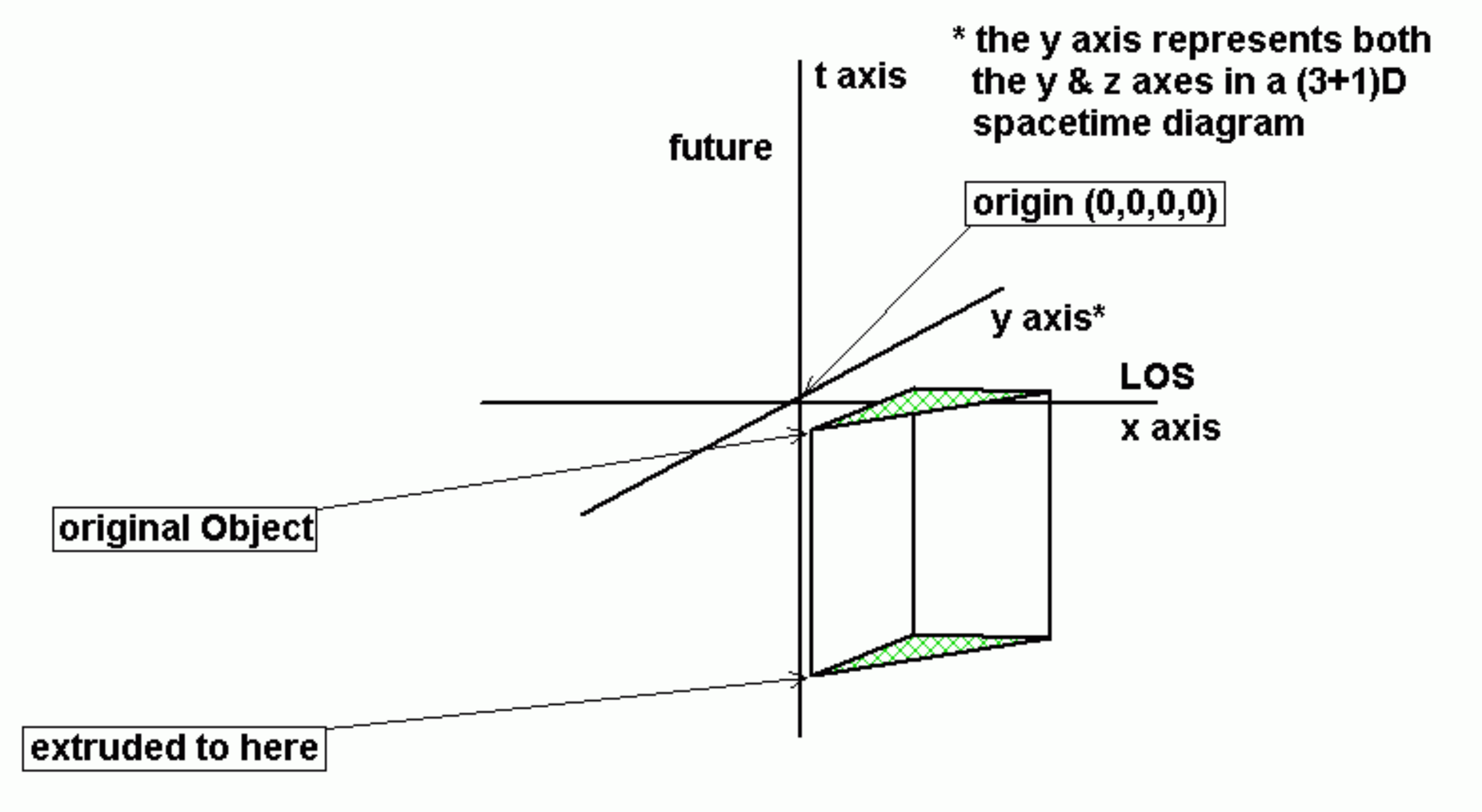}
  \caption{\label{fig:ajp_ob2a} Temporal extrusion: Triangle at rest extruded into prism}
\end{figure}

The object now lies embedded in the XYZ hyperplane that is
orthogonal to the $t$ axis at $t_0$ (\emph{original Object} in
Figure~\ref{fig:ajp_ob2a}). Each of these triangles $f$, and hence
the object composed from them, can be extruded into the 4th
dimension by duplicating the vertices of the triangles with lesser
(or greater) values for the $t$ components. If the object is at rest
in the camera frame, a constant $\Delta t$ can be added to the $t$
component of each of the object's original triangles in the $t_{0}$
hypersurface to create an $f'$ duplicate triangle to be used as the
object's position in the $t_{0}+\Delta t$ hypersurface.

\vspace {-0.2in} \vspace {-0.1in}
\begin{equation}
\label{eqn:one}
       (t_{1}, x_{i}, y_{i}, z_{i})_{f'}~=~(t_{0}, x_{i}, y_{i},
       z_{i})_{f} + (\Delta t, 0, 0, 0)
\end{equation}
\vspace {-0.2in} \vspace {-0.1in}

Where $f=\{1..n\}$ refers to each of the original triangles,
$f'=n+\{1..n\}$ to each of the corresponding extruded triangles, and
$i=\{1,2,3\}$ to each of the corresponding vertices that define each
triangle pair.

As shown in Figure~\ref{fig:ajp_ob2a} where $\Delta t<0$, connecting
the three vertices ($i={1,2,3}$ in Eqn~\ref{eqn:one}) of the
original triangle $f$ with the respective vertices of the extruded
triangle $f'$ creates a 3D prism from the original triangle. Thus
the triangle $f$ exists only between $t_{0}$ and $t_{1}$, inclusive.

The prisms are then tessellated~\cite{tessellated}
into three adjacent tetrahedra as shown in Figure~\ref{fig:fig322}e.
The 3D simplices are necessary for the barycentric algorithm
(described below) used to determine where, on the 3-manifold surface
of the 4D object the intersection with the lightray occurs.


\begin{figure}[htb]
  \centering
  \includegraphics[width=.9\linewidth]{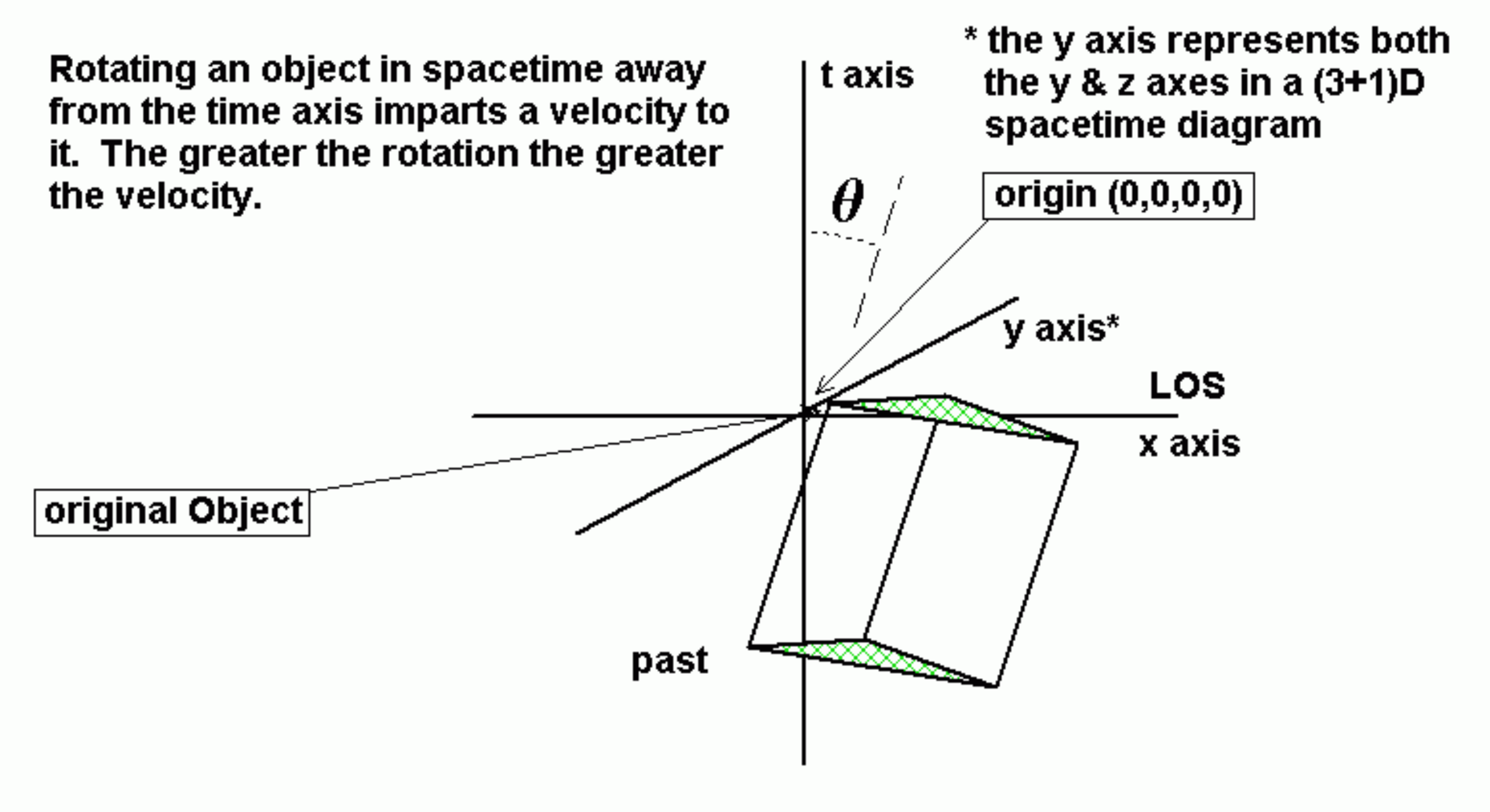}
  \caption{\label{fig:ajp_ob3a} Temporal extrusion not parallel to 't' axis}
  \scriptsize{Object has moved: $velocity = \frac{\Delta x}{\Delta t}$}
\end{figure}

An object's velocity is represented by changing the position of the
extruded end of the triangle (Figure~\ref{fig:ajp_ob3a}) with
respect to the original end:  $x_{end}~=~x_{beg}+\Delta x$ spatial
units. The speed in the camera frame would thus be $\frac{\Delta
x}{\Delta t}~\frac{spatial~units}{time~unit}$. Canceling the units
yields the dimensionless fraction $\frac{\Delta x}{\Delta t}$. A
lightray's slope $c~=~\pm 1.0$ is represented by both the diagonal
lines and the surface of the lightcone of Figure~\ref{fig:Mink2D}.
For the general 3D case, where the distance traveled in time $\Delta
t$ is $\Delta d~=~\sqrt{{\Delta x}^2 + {\Delta y}^2 + {\Delta
z}^2}$, the speed would be $\frac{\Delta d}{\Delta t}$, and
Eqn~\ref{eqn:one} would become:

\vspace {-0.2in} \vspace {-0.1in}
\begin{equation}
\label{eqn:two}
       (t_{1}, x_{i}, y_{i}, z_{i})_{f'}~=~(t_{0}, x_{i}, y_{i},
       z_{i})_{f} + (\Delta t, \Delta x, \Delta y, \Delta z)
\end{equation}
\vspace {-0.2in} \vspace {-0.1in}


\subsection{Viewing 3D Objects in (3+1)D Spacetime}
\label{sec:view}

\begin{figure}[htb]
  \centering
  \includegraphics[width=.9\linewidth]{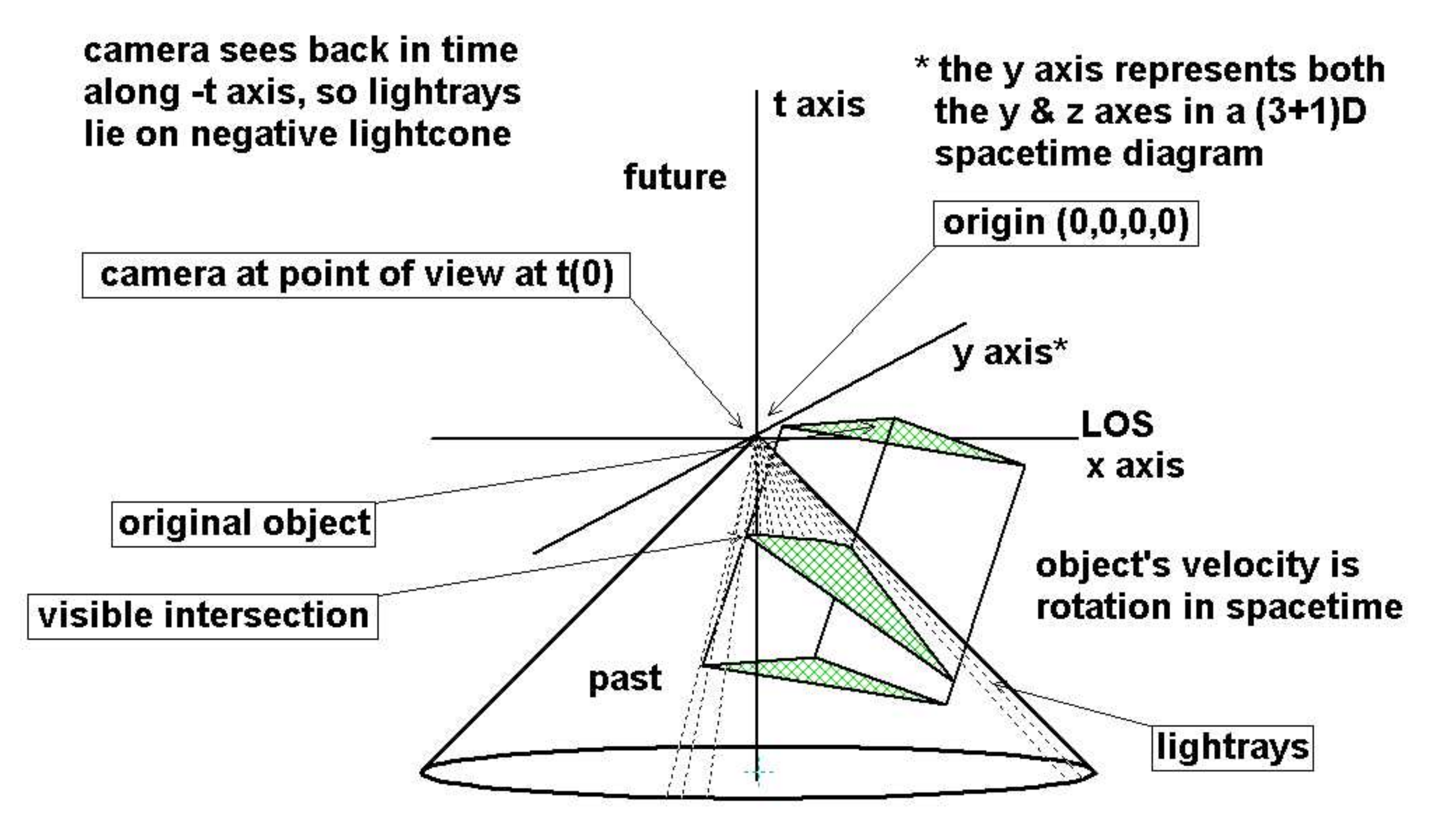}
  \caption{\label{fig:ajp_obj3} Temporal extrusion of moving object with lightcone}
  \scriptsize{A non-relativistic spacetime: finite lightspeed but no Lorentz transform}
\end{figure}

Consider a camera at the origin, whose line-of-sight (LOS) is
collinear with the $x$ axis. Since a lightray's worldline as
depicted in the spacetime diagram lies on the lightcone, an object
must cross the lightcone in the diagram in order to be visible to
the camera.  In fact, the object is visible to the camera only while
it is intersecting that lightcone whose apex is coincident with the
camera (assuming the camera is pointing at the object) as shown in
Figure~\ref{fig:ajp_obj3}.

Figure~\ref{fig:ajp_obj3} depicts a right circular hypercone in
4-space, whose symmetric axis is collinear with the $-t$ axis, and
whose apex is coincident with the camera at the origin $(0,0,0,0)$.
This hypercone's hypersurface, depicted by the inverted cone, has 3
dimensions, sufficient to contain the camera's focal point and the
lightrays entering its lens. Although a 3-manifold in 4-space, this
hypercone is known as a \textbf{\textsl{lightcone}}. A lightcone is
thus the locus of all points that satisfy: \vspace {-0.1in}
\begin{equation}
\label{eqn:dt} (t_{p}, x_{p}, y_{p}, z_{p}) =
(-\sqrt{x_{p}^{2}+y_{p}^{2}+z_{p}^{2}}, x_{p}, y_{p}, z_{p})
\end{equation}

Note that the light travels from the object to the lightcone's apex
at the origin. As depicted by the broken lines representing
lightrays in Figure~\ref{fig:ajp_obj3}, a camera located at the apex
in this 4D model can see only those 3D objects whose extruded
triangles (tetrahedra triads) intersect the lightcone.  The only
visible objects are those with vertex extrusion pairs $(t_{0},
x_{i}, y_{i}, z_{i})$ of the original object and $(t_{1}, x_{i},
y_{i}, z_{i})$ of its extruded end-cap, where

\vspace {-0.2in}
\begin{equation}
\label{eq:t0t1}
           t_{0} \geq \sqrt{x_i^2+y_i^2+z_i^2} \geq t_{1}, \forall~~\{(t_{0}, x_{i}, y_{i}, z_{i})~~\&~~(t_{1}, x_{i}, y_{i}, z_{i})\}
\end{equation}

The intersecting portion of the extruded triangle is depicted by the
triangle labeled \emph{visible intersection} in
Figure~\ref{fig:ajp_obj3}. Note that geometric distortion in the
object is caused by the intersection of the triangle and the
lightcone. An object in the lightcone is easily detected since a
straight line can be intersected with a 3D object in Euclidean
4-space in the same manner as a straight line is intersected with a
2D object in 3-space.


\subsection{Animating Spacetime Objects}

There is no mathematical or geometric limit to an object's speed in
the model, its velocity being the slope of the temporal extrusion
vector. For real physical objects, some physical mechanism must
accelerate the object to the speed with which the object enters the
model's laboratory inertial frame.  We can assume with some
confidence that this speed must be less than that of light. The
physical objects will then maintain an extrusion vector with a
$\frac{|\Delta x|}{\Delta t}$ slope of less than 1.0, or an angle of
less than $45^{\circ}$ with respect to the $t$ axis on the Minkowski
diagram as shown by $\theta$ in Figure~\ref{fig:ajp_ob3a}. Since we
are considering only uniformly moving objects, we can ignore the
specifics of the spacetime rotation that yield the extrusion
angle.~\cite{homogeneity}



\begin{figure}[htb]
  \centering
  \includegraphics[width=.9\linewidth]{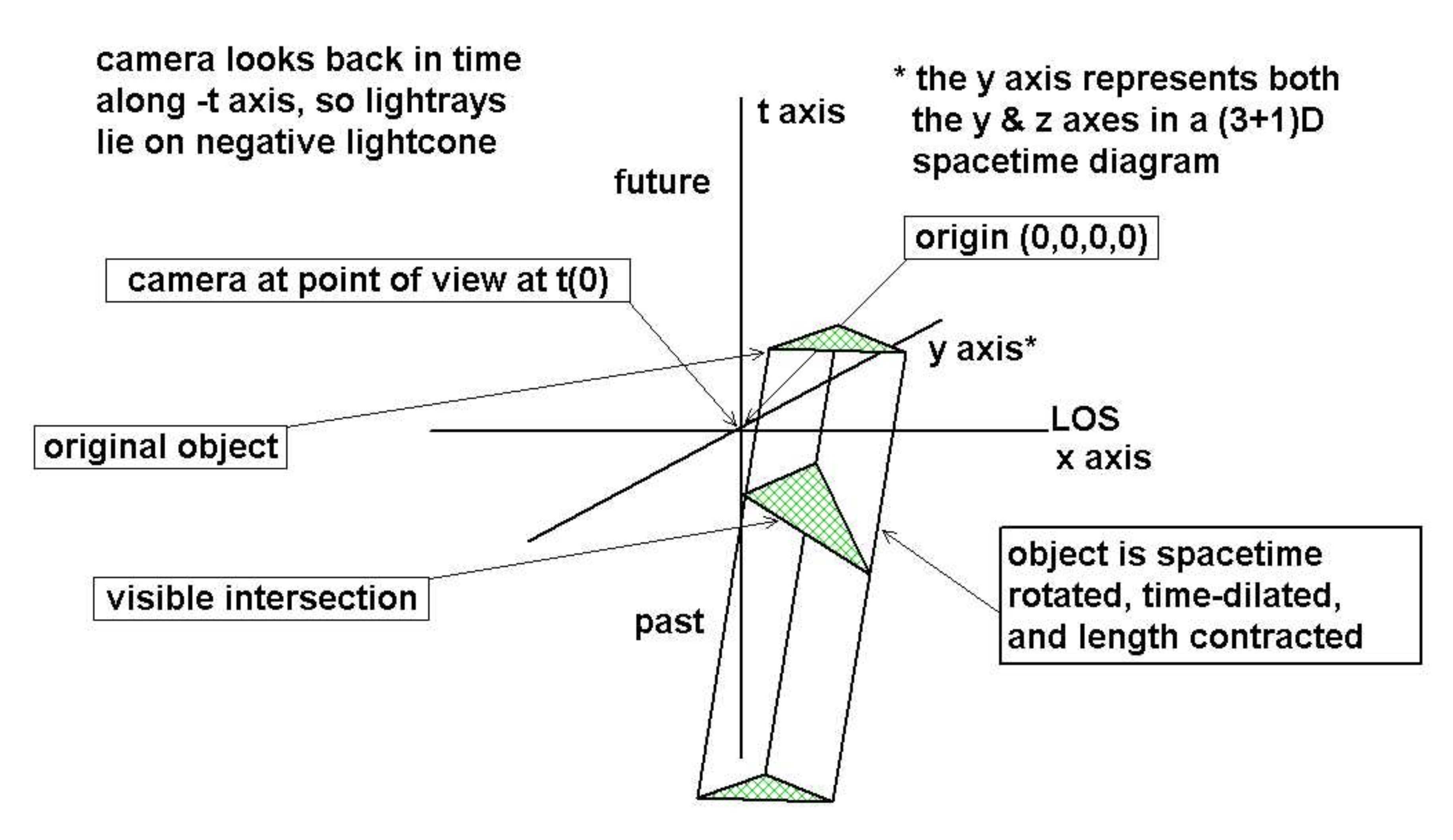}
  \caption{\label{fig:ajp_obj4} Lorentz transformed object}
  \scriptsize{A velocity in the neighborhood of 86.6\% of $c$ yields a $\gamma$ factor around 2.
   The prism is shown length contracted by $\sim\frac{1}{2}$ and its proper time axis is dilated by $\sim2$}
\end{figure}

Two classes of 4D objects have been implemented: one for the finite
lightspeed objects and one for relativistic objects. The first is
inserted into the scene without length contraction or time dilation
as shown in Figure~\ref{fig:ajp_obj3}, and the second is inserted
with the Lorentz transformation as shown in
Figure~\ref{fig:ajp_obj4}. Conceptually, the former may be
considered to have been measured in the laboratory inertial
reference frame's subjective units (it was already length-contracted
and time-dilated), while in the latter case the object was measured
in its own rest frame. The relativistic objects therefore must be
length contracted and time dilated prior to insertion.

The animation procedure is straight forward. For example, to
generate 20 seconds of animation at 10 frames per second ($\Delta t
= 0.1 seconds$), the procedure is as follows.
\begin{enumerate}
\vspace {-0.1in}
\item Beginning with the camera at ($t_0,x_0,y_0,z_0$), a view is rendered and saved.

\item The camera is moved forward in time to ($t_i,x_0,y_0,z_0$), where $t_i=t_{i-1}+\Delta t$ and the view is rendered and saved;
\label{sec:repeat}

\item Repeat from step \ref{sec:repeat} while $t_i<20$.
\end{enumerate}

Notice that the camera's spatial components ($x,y,z$) do not change,
only the time component of the camera position.
Crucial to the simplicity of the procedure is the fact that the 4D
object's bounding surfaces (and the tessellating tetrahedra that
comprise those surfaces) \textbf{\emph{do not change}}. The 4D world
is \textbf{\emph{static}}. Only the point of intersection of the
lightcone changes as the camera and its lightcone progress along the
$t$ axis.


\subsection{4D Intersection Algorithm}

Lightcone crossing events are detected by solving for the
intersection of a lightray with each of an object's bounding
tetrahedra. The set of lightrays is defined as that set of 4D
straight lines passing from the camera through each of the pixels in
the viewplane's pixel grid and out into 4D space. Using a 4D
implementation of the barycentric algorithm to compute the
intersections of the ray with all tetrahedra faces, we select the
intersecting event nearest to the camera (with the $t$ value closest
to $0.0$). The array of 1D lightrays that originate from the gridded
viewplane results in a 2D image of the objects projected onto that
viewplane.

Since the objects have been Lorentz transformed prior to the
intersection, such that their geometry is correct for the camera
frame in which the intersection occurs, the geometric components of
the lighting model, the surface normal and the reflection angle, can
be used to approximate the pixel shade just as with a conventional
lighting model in 3D rendering.

Photorealistic rendering requires the addition of lighting effects
such as Doppler shift\cite{Hsiung:1990b} and the searchlight effect,
which could dominate the rendered image and obscure the
visualization of the object's geometry.\cite{Weiskopf:1999b}  For
this reason, these effects were not implemented in this model.

\section{Results}
\label{sec:resu}

\begin{figure}[p]
  \centering
  \includegraphics[width=0.40\linewidth]{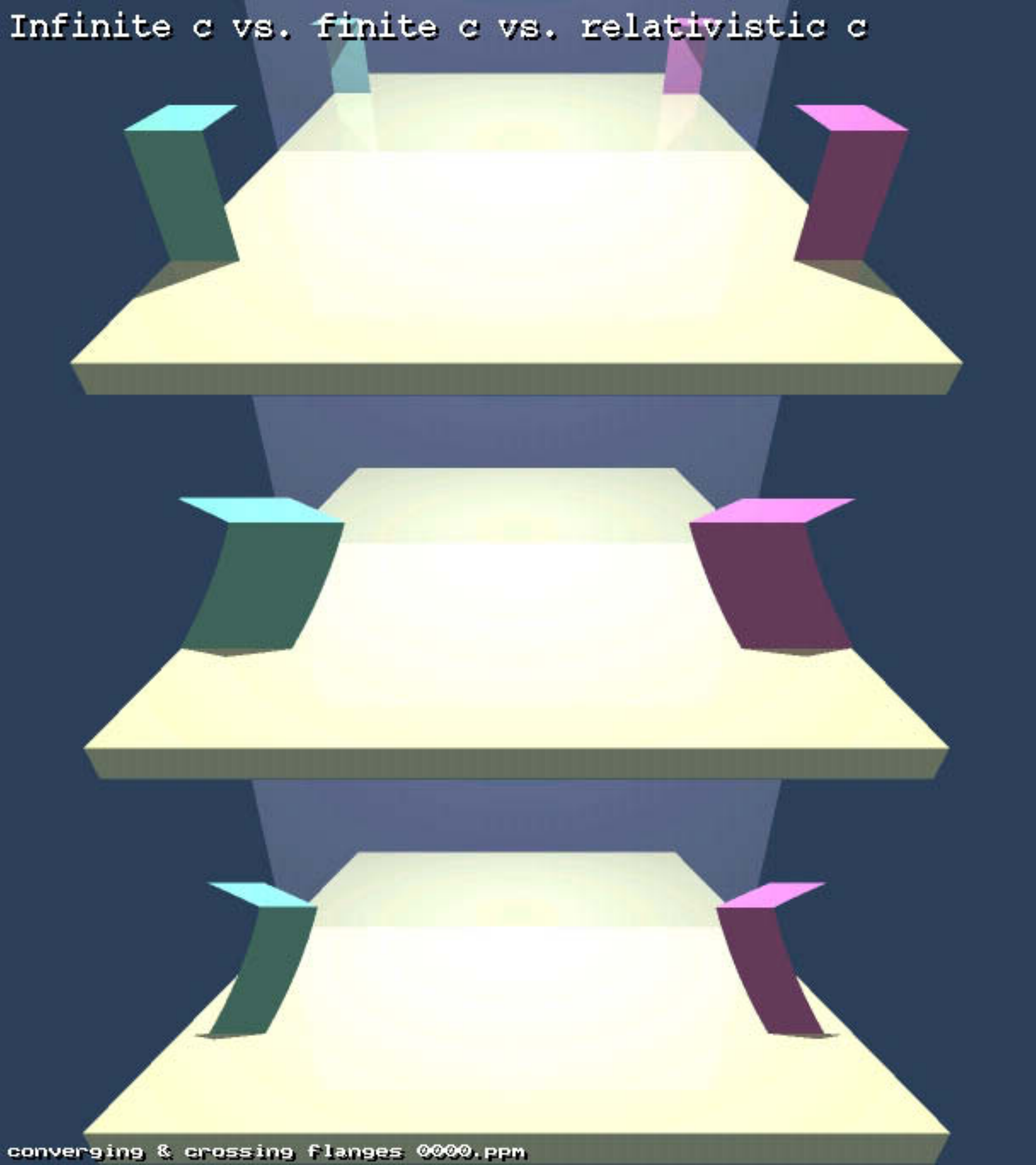}
  \includegraphics[width=0.40\linewidth]{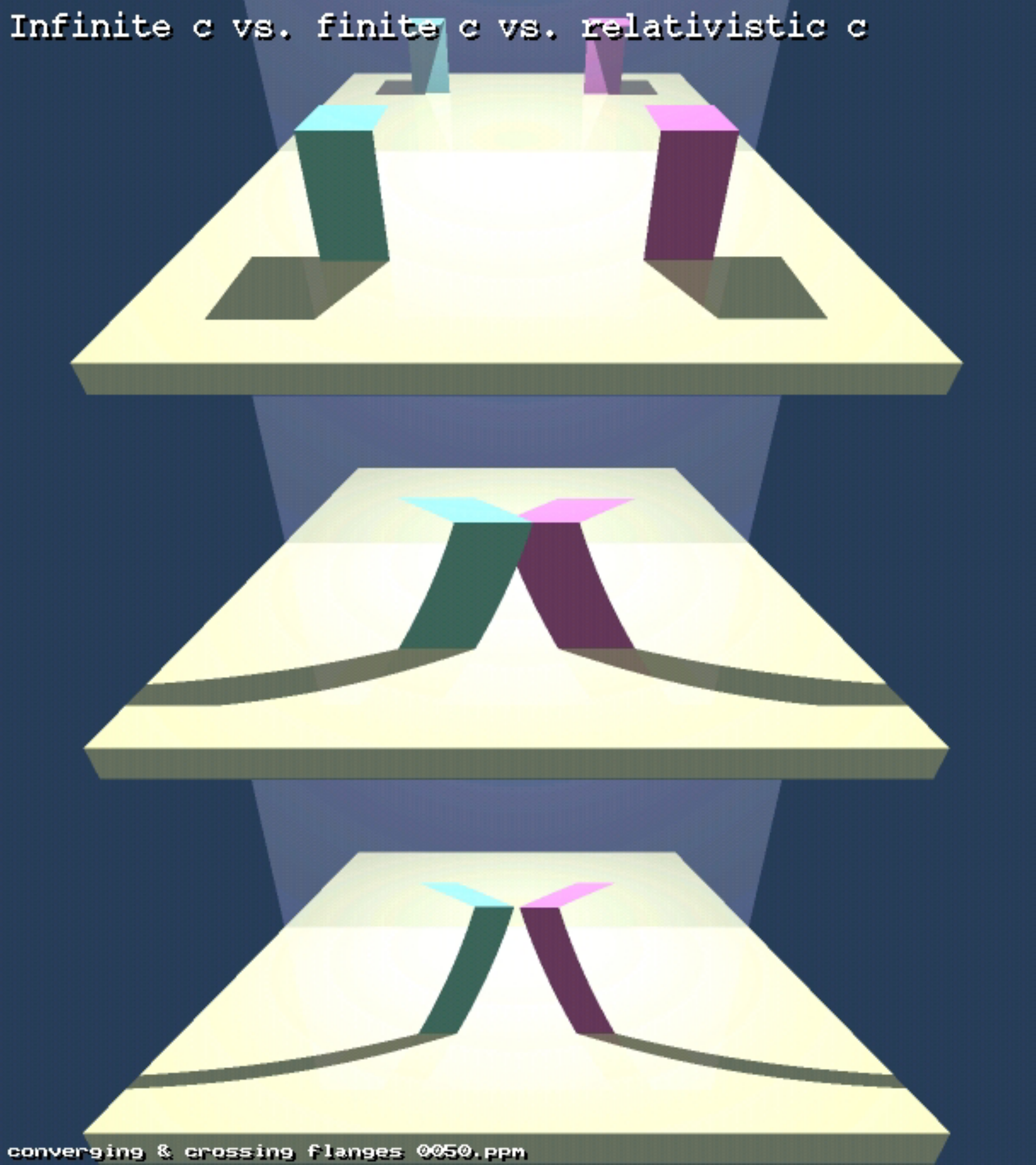} \\
  \includegraphics[width=0.40\linewidth]{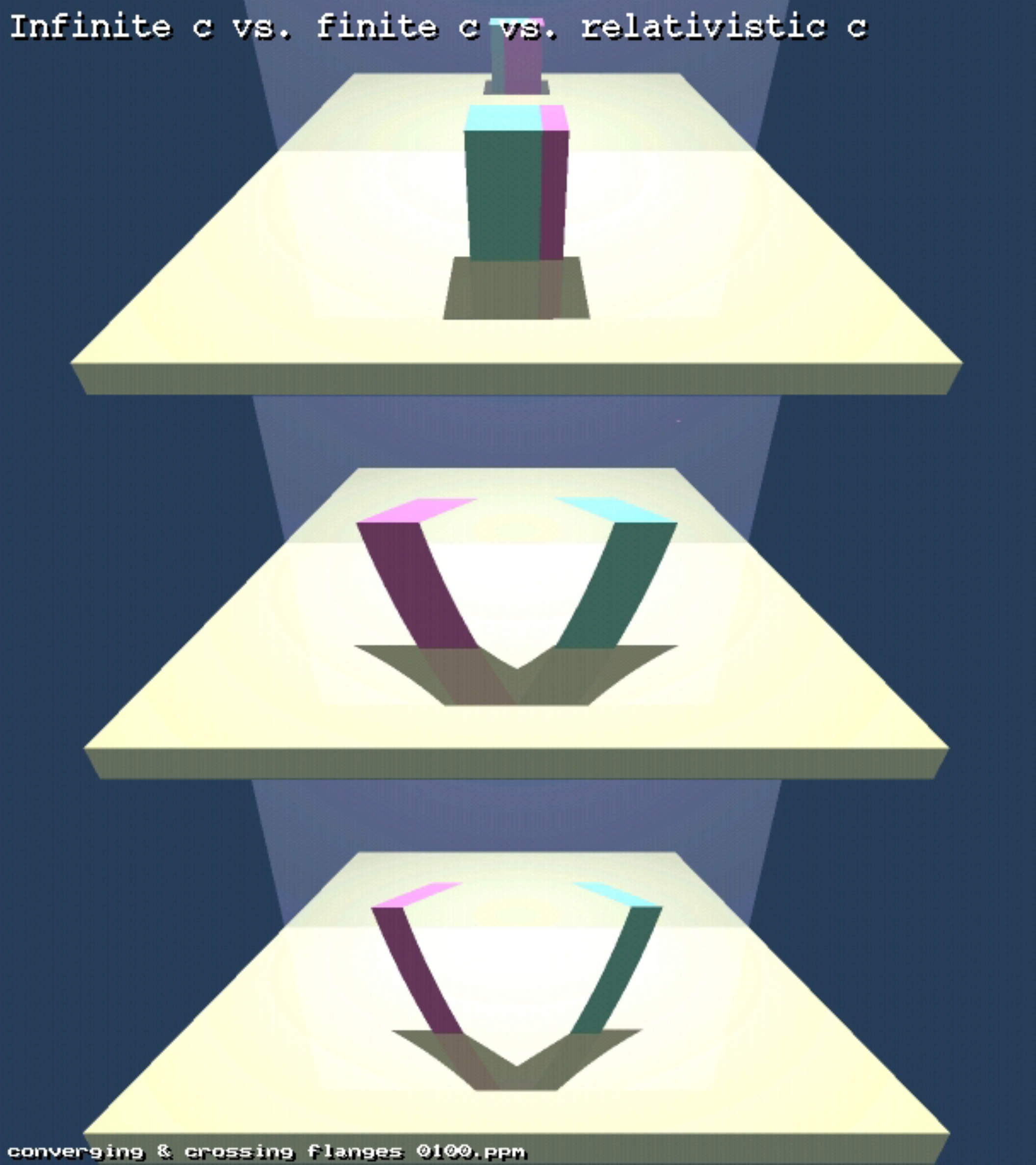} \\
  \includegraphics[width=0.40\linewidth]{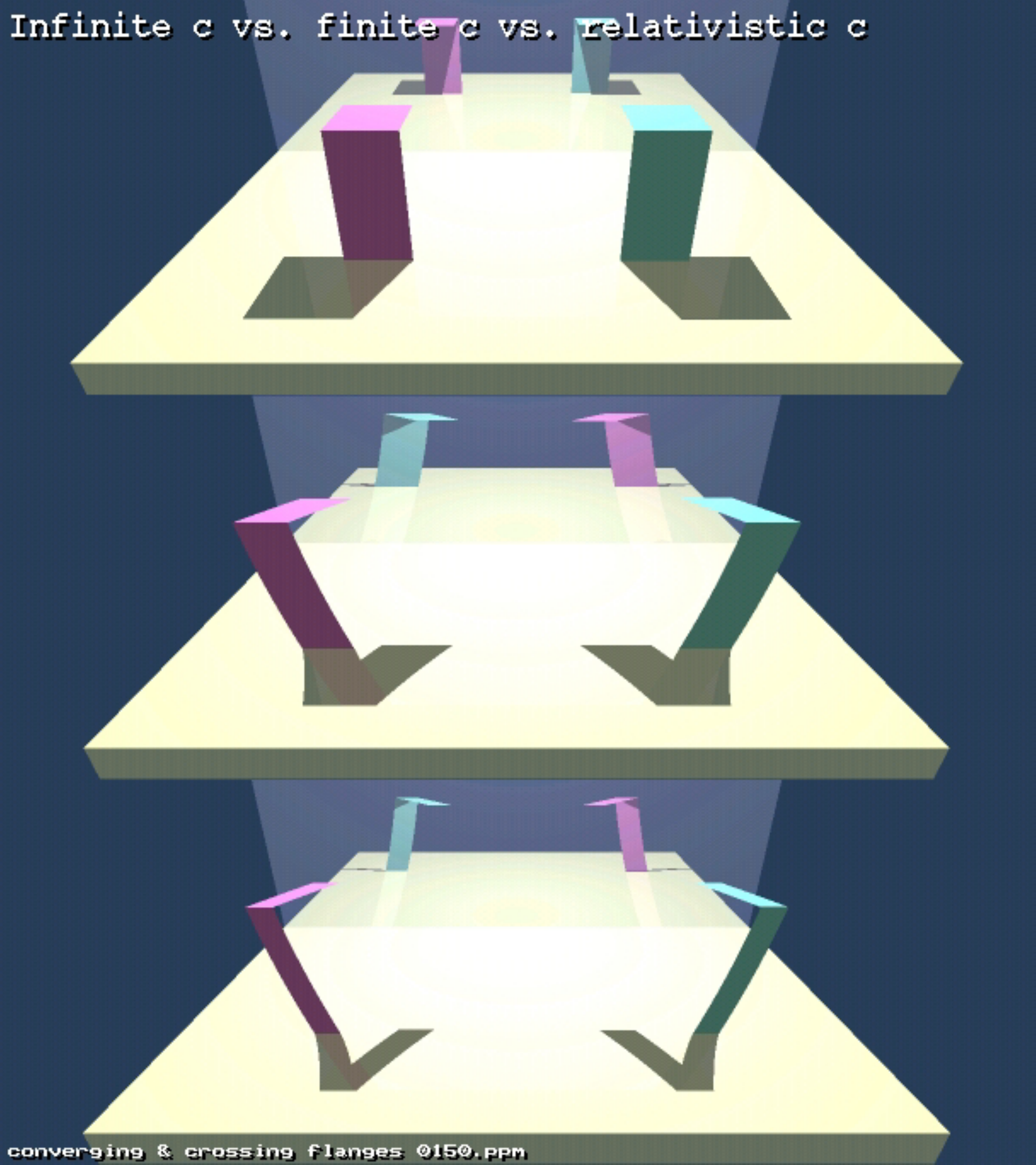}
  \includegraphics[width=0.40\linewidth]{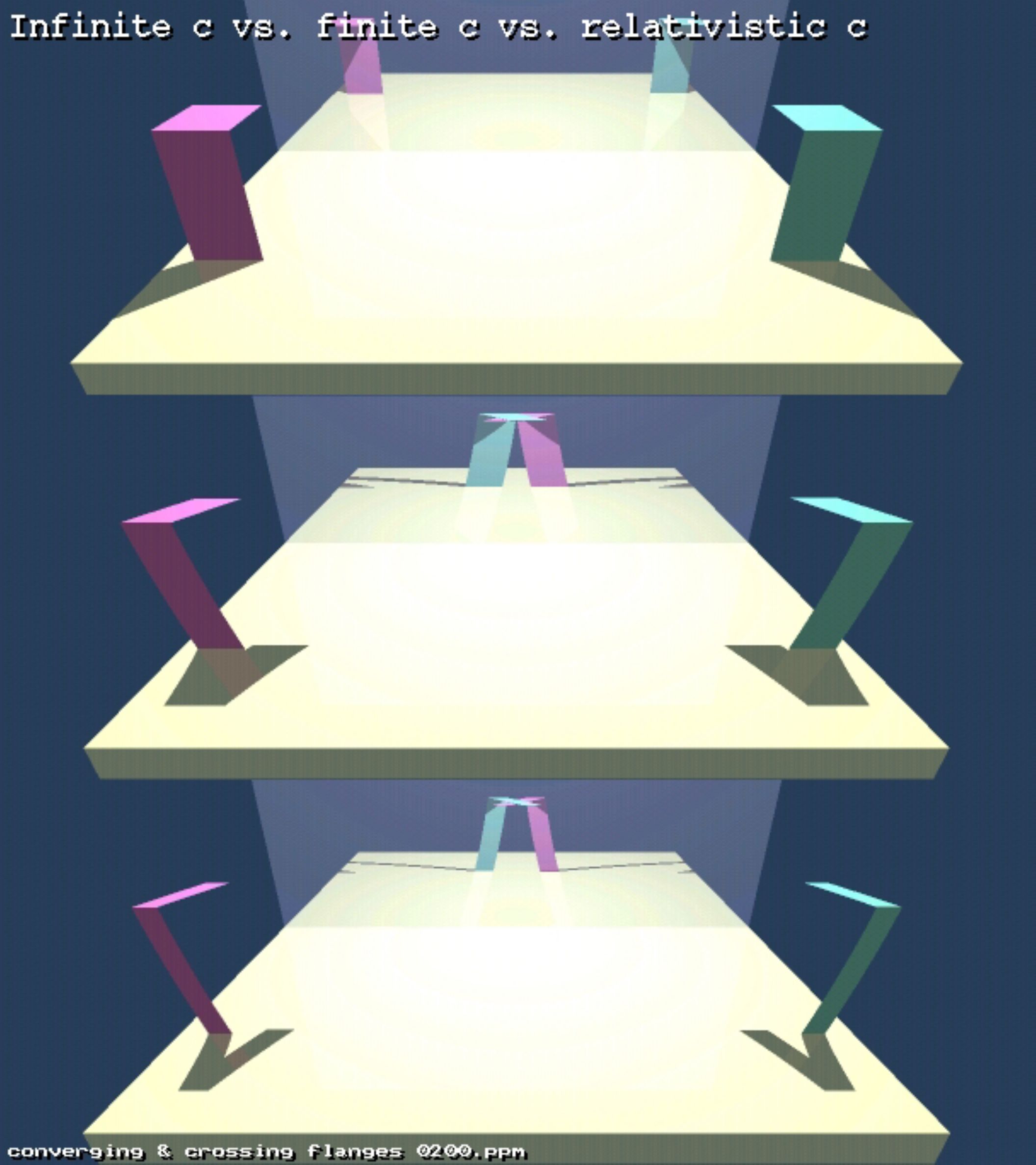}
  \caption{\label{fig:figteas}\scriptsize{Sequential images of two 4D objects converging then crossing at $0.866c$ on a mirrored
  background}}
  \scriptsize{Top row: infinite lightspeed 4D raytracing; \\
   Center row: pre-relativistic spacetime model (no Lorentz transform; \\
   Bottom row: relativistic spacetime model (with Lorentz transform}
\end{figure}

Three models of relativistic motion are displayed in the sequential
images in Figure~\ref{fig:figteas}. The object displayed is a flange
(angle bracket) 2 light-seconds wide by 2 light-seconds deep by 4
light-seconds tall. Its thickness is negligible (being constructed
of four 2D triangles). The top row shows the traditional ray-tracing
technique, where the lightspeed is effectively infinite. In the
middle row, the pre-relativistic optical effects are shown, while in
the bottom row, the relativistic effects are displayed. The finite
lightspeed camera (top row) was moved ahead in time 18.675 seconds,
an amount equal to the lightspeed delay from the center of the stage
to the camera, so that the flanges appear to be in approximately the
same positions.

The scene is set upon a stage with an overhead light source, both at
rest in the camera frame. Two flanges approach, cross, and depart
the centerline of the scene at $0.866 c$. The geometric distortions
of the center row are due exclusively to classical aberration. Those
of the third row are due to relativistic aberration.

The stage's mirrored backdrop shows the reflections of the flanges
from behind. Note the difference in the positions of the reflections
in the three rows. The top row shows the instantaneous reflections
of the flanges, while the middle and bottom rows show the retarded
reflections due to the lightspeed delay imposed by the added
distance to be traveled by the lightray from the object to the
mirror and back. The distances modeled are on the order of the size
of the Jovian system.~\cite{Jovian}


Note the bottom flanges \emph{appear} to cross each other before the
top flanges. Note also, that even with this head start, the top
flanges arrive at their respective edges at the same time as the
bottom flanges. The bottom flanges appear to approach faster and
retreat slower than the top flanges. This is the visual evidence of
the pre-relativistic optical effect known as classical aberration.
The flanges approaching the centerline of the stage are obliquely
approaching the camera. Aberration causes the angle from the
centerline to the flanges to appear smaller than the proper angle of
incidence, resulting in the object appearing closer to the
centerline, or ahead of the object's proper position as depicted in
the top view.

This is true for both the leading and the trailing edges of the
flange, independently. As a result, the leading edge, which is
closer to the centerline, appears to have moved further than the
trailing edge, giving the impression of a wider flange. The opposite
effect occurs as the flanges move away from the centerline. The
flanges appear to incrementally speed up and simultaneously contract
as they move relativistically away from the camera. These aberration
effects are apparent in the bottom two images of
Figure~\ref{fig:figteas}.

\section{Conclusions}

This paper has visually demonstrated that implementing a simple
algorithm that consists of a finite lightspeed component and a
length contraction component yields special relativistic
visualizations similar to those using more complex visualization
systems.\cite{Weiskopf:2005} We have viewed the difference between
pre-relativistic optical effects due to a finite lightspeed and
those effects predicted by special relativity.

We have demonstrated that 3D animated sequences can be generated
from a static 4D Euclidean spacetime. Furthermore, we have
demonstrated that 3-space can be visualized as the intersection of a
lightcone and Euclidean 4-space, where the slope of the lightcone's
hypersurface determines the constant lightspeed in the model. The
model has been described using only algebra and Euclidean geometry.

\section*{Acknowledgements}

We would like to thank Dr. Daniel Weiskopf and Dr. Arvind Rajaraman
for their invaluable comments and suggestions. A special thanks also
to Dr. James Arvo for his ToyTracer raytrace kernel code.

\bibliographystyle{unsrt}

\begin{thebibliography}{10}
\bibitem{Jovian}
At the recorded animation rate (your playback may differ), the size
of the
  stage is about 6 million Km on a side (20 light-seconds), easily large enough
  to encompass the planet Jupiter and the orbits of its four largest moons.
\end{thebibliography}

\end{document}